\documentclass[twocolumn,amsmath,amssymb,aps]{revtex4}
\usepackage{graphicx}
\usepackage{color}
\usepackage{natbib}
\usepackage{xcolor}
\usepackage{comment}
\usepackage{mathtools}

\usepackage[colorlinks=false, linkbordercolor=red, citebordercolor=green, urlbordercolor=cyan, pdfborderstyle={/S/U/W 1}]{hyperref}
\usepackage{soul}

\begin{document}

\title{Unconditional accumulation of nonclassicality in a single-atom mechanical oscillator}
\author{L.~Podhora$^{1}$}
\author{T.~Pham$^{2}$}
\author{A.~Le\v{s}und\'{a}k$^{2}$}
\author{P.~Ob\v{s}il$^{1}$}
\author{M.~\v{C}\'{i}\v{z}ek$^{2}$}
\author{O.~\v{C}\'{i}p$^{2}$}
\author{P.~Marek$^{1}$}
\author{L.~Slodi\v{c}ka$^{1}$}
\email{slodicka@optics.upol.cz}
\author{R.~Filip$^1$}

\affiliation{$^1$ Department of Optics, Palack\'{y} University, 17. listopadu 12, 771 46 Olomouc, Czech Republic \\
$^2$ Institute of Scientific Instruments of the Czech Academy of Sciences, Kr\'{a}lovopolsk\'{a} 147, 612 64 Brno, Czech Republic}

\date{\today}

\begin{abstract}

We report on the robust experimental accumulation of nonclassicallity of motion of a single trapped ion. The nonclassicality stems from deterministic incoherent modulation of thermal phonon number distribution implemented by a laser excitation of nonlinear coupling between the ion's internal - electronic levels and external - motional states. We demonstrate that the repetitive application of this nonlinear process monotonically accumulates the observable state nonclassicality. The output states converge to a phonon number distribution with high overlap with a particular Fock state and visible quantum non-Gaussian aspects including corresponding negative Wigner function. The resulting oscillator states prove deterministic transition in the hierarchy of quantum non-Gaussianity up to four phonons. This transition is very robust against experimental imperfections and produces increasing entanglement potential.

\end{abstract}

\pacs{}

\maketitle

\section{Introduction}

Quantum nonclassical states represent a directly observable product of nonlinear quantum processes and are a paramount resource for studies of light and matter~\cite{glauber1963coherent,sudarshan1963equivalence,davidovich1996sub,raimond2001manipulating,leibfried2003quantum,xiang2013hybrid}, processing of quantum information~\cite{braunstein2005quantum,devoret2013superconducting,andersen2015hybrid}, and have proven to be beneficial for a broad range of metrological and sensing applications~\cite{pezze2018quantum,rivas2010precision,chalopin2018quantum,taylor2016quantum,tan2019nonclassical}. One of the obstacles for their full utilization in mechanical oscillators is often severely limited effective creation probability for optical or microwave measurements and still large energy of thermal environment for solid-state experiments at cryogenic temperatures. In a vast majority of experimental demonstrations, generation of nonclassical phonon-number states utilizes an initial ground state of the system with very low entropy combined with strong nonlinear interaction, high quality projection measurement, or both~\cite{meekhof1996generation,lvovsky2001quantum,ourjoumtsev2006quantum,deleglise2008reconstruction,hofheinz2008generation,yukawa2013generating,kienzler2017quantum,chu2018creation,wolf2019motional}. In many mechanical systems the available level of control is mostly insufficient for the initial step of entropy minimization and, at the same time, nonlinear couplings are weak to be employed on the relevant timescales and efficient noiseless projective measurement is not available.

To overcome these natural limitations, we report on the experimental demonstration of unconditional accumulation of nonclassicality for a single-atom mechanical oscillator prepared initially in thermal state with energy of several motional quanta by employing the scheme presented in the seminal paper by~\emph{R. Blatt et al.}~\cite{blatt1995trapping}. We use the basic Rabi interaction corresponding to a coupling between the mechanical oscillator states of a single atom and its electronic spin state~\cite{rabi1936process,kockum2019ultrastrong,forn2019ultrastrong,chang2018colloquium}. Such nonlinear interaction is thus applicable to a broad range of experimental systems which allow for a direct implementation of a blue detuned Rabi interaction, including trapped atoms, superconducting qubits coupled to a microwave radiation, or increasing variety of solid-state systems and optomechanical platforms~\cite{ding2017quantum,fluhmann2019encoding,campagne2019stabilized,niemczyk2010circuit,lo2015spin,todorov2010ultrastrong,hartke2018microwave,kounalakis2019synthesizing}. For oscillator's thermal state and ground state electronic spin the Rabi coupling already realizes a complex deterministic modulation of a phonon number distribution. This modulation brings nonclassicality in phonon-number distribution of the oscillator as a counterpart of a fundamental collapse and revival effect in the electronic spin~\cite{rempe1987observation,meunier2005rabi,hofheinz2008generation,braumuller2017analog,assemat2019quantum}. We experimentally test an accumulation of the nonclassicality through repetitive application of the blue-detuned Rabi
interaction with reinitialized electronic state which results in a monotonous accumulation of the generated nonclassicality manifested in several witnesses of nonclassicality. For any temperature of the oscillator, accumulated nonclassicality becomes apparent in the phonon number distribution converging to a dominant Fock states with corresponding negative Wigner function and level of quantum non-Gaussian hierarchy~\cite{straka2018quantum}.
The quantum non-Gaussian properties provably enhance for each of the consecutively repeated interactions, while each of these interactions is already theoretically provably sufficient for nonclassicality generation. While the presented nonclassicality accumulation shares some phenomenological similarities with the conventional nonclassicality distillation~\cite{heersink2006distillation,filip2013distillation}, it crucially differs in the fact, that it is unconditional. Moreover, we do not dynamically engineer neither state of the atoms nor coupling strength to reach accumulation, as in the reservoir engineering~\cite{poyatos1996quantum,kienzler2015quantum} or in methods exploiting adiabatic passage~\cite{um2016phonon,cirac1994nonclassical}. The accumulation of nonclassical properties can be viewed as quantum non-Gaussian mechanical counterpart of nonlinear optical parametric processes in a cavity where the initial weak nonclassical effect generated by a single implementation of the nonlinear interaction is gradually accumulated~\cite{bachor2004guide}.

\section{Nonlinear interaction with atomic mechanical oscillator}

\begin{figure}[!t]
\begin{center}
\includegraphics[width=1\columnwidth]{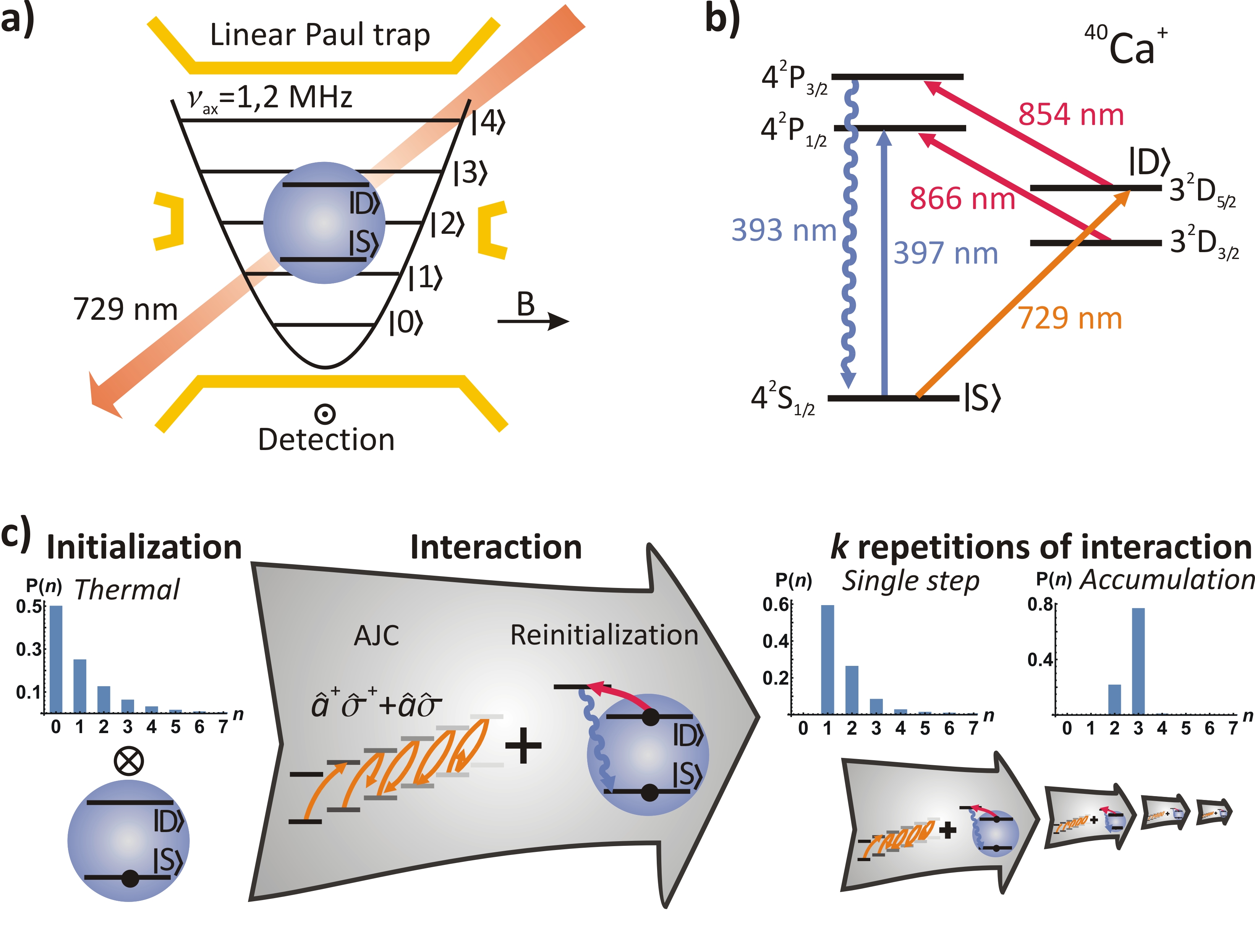}
\caption{A depiction of nonclassicality accumulation in a mechanical oscillations at a frequency of $1,2$ MHz corresponding to a single trapped ion. a) shows the experimental apparatus with employed linear Paul trap, 729~nm laser excitation and scattered light detection geometry. Occupation of phonon-number states $|n>$ is deterministically modulated by a blue-detuned Rabi interaction with a two-level system consisting of electronic states $|S\rangle$ and $|D\rangle$. The relevant internal energy level scheme of $^{40}{\rm Ca}^{+}$ ion is shown in b). A simplified experimental sequence for an unconditional generation of nonclassical states from initially classical thermal population of atomic motion is shown in c). The nonlinear AJC coupling is followed by a reinitialization of internal atomic population to the $|S\rangle$ level. This process is repeated $k$-times.}
\label{fig:setup}
\end{center}
\end{figure}

The presented experimental demonstration of deterministic nonclassicallity accumulation utilizes a high degree of control of $k$-times repeated coupling between the mechanical motion of a single trapped ion and its internal electronic state~\cite{blatt1995trapping}. The experimental setup comprises of a linear Paul trap for spatial localization of single $^{40}$Ca$^+$ ion. A simplified excitation geometry is shown in the Fig.~\ref{fig:setup}-a). To realize the deterministic nonclassicality accumulation, we employ ion's axial motion with secular frequency set to $\omega_{\rm ax}=2 \pi\times 1.2$~MHz. Excitation laser beams at 397~nm, 866~nm and 729~nm are propagating under angle 45~degrees with respect to trap axial direction. The 397~nm optical pumping and 854~nm repumping beams are propagating along the direction of the applied magnetic field $\overrightarrow{B}$ with circular and elliptical polarizations, respectively. 
The experimental sequence begins with a 1~ms period of Doppler cooling using the 397~nm laser and the 866~nm beam is used for reshuffling the atomic population from the metastable 3D$_{3/2}$ manifold, see Fig.~\ref{fig:setup}-b) for the relevant energy level scheme. In the next step, the optical pumping pulse prepares the atomic population in the 4S$_{1/2}(m=-1/2)$ Zeeman sub-level. The energy of the initial thermal motional state is set by controlling the length of sideband cooling sequence, which is implemented using the 729~nm laser tuned to the first red motional sideband of the 4S$_{1/2}(m=-1/2)\leftrightarrow 3{\rm D}_{5/2} (m=-5/2)$ transition together with a weak 854~nm beam which reshuffles the 3D$_{5/2}(m=-5/2)$ population to the ground state 4S$_{1/2}(m=-1/2)$.

For any step $k$ of the accumulation, the interaction between ion's axial motional mode with an input state $\rho_{\rm k}=\sum_{n=0}^{\infty} P_{\rm k} (n) |n\rangle\langle n|$ and internal quasi-two level system corresponding to the transition $|S\rangle =4{\rm S}_{1/2}(m=-1/2) \leftrightarrow |D\rangle = 3{\rm D}_{5/2}(m=-1/2)$ is realized by the excitation of the first blue motional sideband using the 729~nm laser. An observable large number of Rabi oscillations with close to full contrast with a Rabi frequency on a blue motional sideband corresponding to $g=2 \pi\times 5.8$~kHz in our setup allows for implementation of coherent blue-detuned Rabi (anti-Jaynes-Cummings) interaction between spin and motion with high fidelity, see Supporting information~\ref{subsec:processing}. However we note, that successful implementation of the nonclassicality accumulation only requires small coherence of the anti-Jaynes-Cummings (AJC) interaction and is thus applicable to systems with much smaller coherence of nonlinear coupling and with higher presence of noise. 
The accessible high control of trapped ion states should be viewed as a feasible tool for the realization of proof of principle characterization of the scheme rather than demanding condition.

In the Lamb-Dicke regime, the AJC interaction corresponding to the excitation of a blue motional sideband of trapped ions can be well approximated by effective Hamiltonian~\cite{blatt1995trapping,leibfried2003quantum}
\begin{equation}
\hat{H}_{\rm blue}=g/2 (\hat{a}^{\dag} \hat{\sigma}_+ + \hat{a} \hat{\sigma}_-).
\label{eq:blueInt}
\end{equation}
where $g$ is the coupling strength, $\hat{a}$ is a bosonic operator acting on the axial harmonic motional mode, and $\hat{\sigma}_{+}, \hat{\sigma}_{-}$ are two-level raising and lowering operators of electronic spin, respectively.
For oscillator with input phonon number distribution $P_{\rm k}(n)$ in $k$-th step of the procedure and the electronic spin prepared in the ground state $|S\rangle$, the AJC interaction results in the state with modulated phonon populations,
\begin{eqnarray}
\rho_{\mathrm{k+1}}=\sum_{n=0}^{\infty}P_{\rm k}(n)\left[\cos\left(gt/2\sqrt{n+1}\right)\right]^2|n\rangle\langle n|+\nonumber\\
\sum_{n=0}^{\infty}P_{\rm k}(n)\left[\sin\left(gt/2\sqrt{n+1}\right)\right]^2|n+1\rangle\langle
n+1|,
\label{eq:singleStepIdeal}
\end{eqnarray}
where $t$ depends on the laser excitation time and $gt$ therefore stands for the effective area of the driving pulse.
The output state $\rho_{\rm k+1}=\sum_{n=0}^{\infty} P_{\rm k+1}(n) |n\rangle\langle n|$ is fully defined by its phonon population $P_{k+1}(n)$ and is separated from the state of the electronic spin. It can be therefore directly used in the next iteration step, which is again represented by (\ref{eq:singleStepIdeal}). Note that the accumulation process is the same in all $k$ steps; the pulse area $gt$ is not changed.

Already for an initial thermal state with $P_0(n)=\sum_{n}\langle \hat{n} \rangle^{n}/(\langle \hat{n} \rangle +1)^{n+1}$, the incoherent modulation~(\ref{eq:singleStepIdeal}) can deterministically result in a nonclassical state for a broad range of electronic and motional thermal states and excitation pulse lengths~\cite{blatt1995trapping,slodivcka2016deterministic,marek2016deterministic}.
The nature of the operation depends on the pulse area. For our proof-of-principle demonstration of the concept we choose $gt=\hbar \Omega_{\rm c} \eta t \sim \pi$ and the initial internal state corresponding to $|S\rangle$, because it corresponds to an addition of one phonon to a ground state of the oscillator. Here $\eta_{729}= 0.063$ is the Lamb-Dicke parameter for the interaction with the 729~nm beam, $\Omega_{\rm c}=2 \pi\times (92\pm 1)$~kHz is the Rabi frequency on the corresponding carrier transition and the length of the laser pulse is set to 91~$\mu$s. Beyond initial motional ground state, single step $k=1$ deterministically shifts the phonon number distribution, as depicted at the Fig.~\ref{fig:setup}-c).

We test the nonclassical properties of the shifted phonon-number distributions $P_{\rm 1}(n)$ resulting from the first round of iteration~(\ref{eq:singleStepIdeal}) for input thermal phonon populations $P_{0}(n)$ with various energies. This is followed by reshuffling of the excited $|D\rangle$ level population back to the initial ground state $|S\rangle$ using the excitation by 854~nm laser to the  $4\rm{P}_{3/2}$ manifold followed by the emission of a single 393~nm photon. In addition, a short optical pumping 397~nm pulse ensures that atomic population is pumped to the initial Zeeman sublevel $|S\rangle\sim4{\rm S}_{1/2},m=-1/2$. The reshuffling effectively corresponds to resetting the internal electronic state of the ion and makes the operation unconditional. However, at the same time, the random recoils from the resonant 854~nm laser excitation and 393~nm emission result in a small heating of mechanical populations with the total weight given by the probability of finding the ion in the $|D\rangle$-level after the interaction. In the Lamb-Dicke regime, such redistribution effectively corresponds to the interaction of mechanical mode with a thermal reservoir
and it happens only with the probability of finding the ion in the excited state $|D\rangle$ after the process~(\ref{eq:singleStepIdeal}). A detailed model used for employed process simulations can be found in Supporting Information~\ref{subsec:model}.

\begin{figure}[!t]
\begin{center}
\includegraphics[width=1\columnwidth]{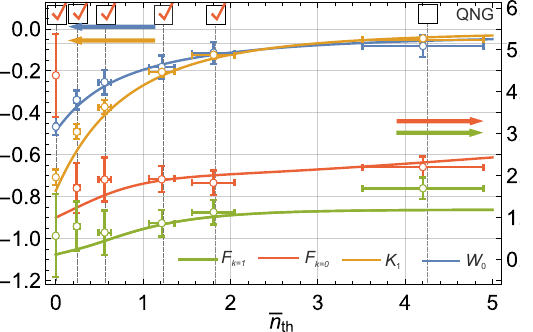}
\caption{The results of evaluation of nonclassicality for the measured phonon number distributions $P_1(n)$ after single nonlinear AJC interaction of atomic mechanical oscillator prepared in thermal state. The Fano factors evaluated for initial and generated phonon populations demonstrate the conversion to sub-Poissonian statistics for states with low initial thermal energy $\overline{n}_{\rm th}$. The evaluated negative Klyshko's criteria $K_1$~\cite{klyshko1996observable} for each output distribution unambiguously confirm a strong nonclassicality of the generated states for a broad range of initial thermal energies $\overline{n}_{\rm th}$. In addition, the observed negative values of the Wigner quasidistribution $W(0,0)$ suggest that the generated state is always non-Gaussian. Moreover, quantum non-Gaussianity criteria (QNG)~\cite{straka2018quantum} show impact of multiphonon contributions. The measures evaluated from the experimental data are displayed as full circles with error bars corresponding to a three standard deviations. The solid lines correspond to theoretical predictions for AJC interaction~(\ref{eq:singleStepIdeal}) with a $gt=\pi$ and for given $\overline{n}_{\rm th}$ with no free fitting parameters.}
\label{fig:singleAd}
\end{center}
\end{figure}

Let us first analyze this single step of the procedure. We have applied nonlinear AJC~(\ref{eq:singleStepIdeal}) with $gt=\pi$ to the oscillators initially in thermal distribution $P_0(n)$ with different mean energies. The phonon number distributions $P_1(n)$ were then obtained from fits of measured Rabi oscillations on the blue motional sideband. The initial states after sideband cooling correspond very well to the states with close to ideal Bose-Einstein statistics within the errors evaluated using the Monte-Carlo simulations with input uncertainties corresponding solely to the projection noise in measured of the qubit state. The phonon number distributions were then used to evaluate nonclassicality of the produced states. Fig.~\ref{fig:singleAd} shows results of the evaluation of Fano factor $F=\langle (\Delta n)^2\rangle/\langle n\rangle$, Klyshko's criteria for nonclassicality~\cite{klyshko1996observable}, and values of Wigner function at the center of the phase space $W(0,0)$.
These results confirm that AJC~interaction can be used for deterministic generation of nonclassicality for a broad range of initial thermal energies of the atomic motion. The nonclassicality criteria evaluated for the measured $P_1(n)$ are in very good qualitative agreement with simulation which has been used without any fitting parameters. Negative Wigner function $W(0,0)<0$ proves quantum non-Gaussianity for all measured data. However, multiphonon contributions are already too high for $\bar{n}=4.2\pm 0.2$, therefore, quantum non-Gaussianity is not sufficient from perspective of the hierarchy~\cite{straka2018quantum}. Satisfying the $n$-th member of the non-Gaussian criteria hierarchy signifies that there cannot exist any Gaussian state that has the same values $P(n)$ and $\tilde{P}_{n+1} = 1-\sum_{k=0}^n P(k)$ as the observed state. Please see Supplemental material S3 for details. To accumulate this quantum non-Gaussian aspects, we therefore focus on initial thermal state with $\bar{n}=1.19\pm 0.04$. A detailed description of the phonon number distribution measurement and analysed nonclassicality criteria can be found in the Supporting information~\ref{subsec:processing} and~\ref{subsec:criteria}. In addition, the detailed investigation of the initial phonon statistics shows good agreement with the thermal Bose-Einstein distribution and the nonclassicality generation can thus be unambiguously attributed to the implemented deterministic nonlinear interaction. See Supporting information~\ref{subsec:singleStep} for the corresponding full statistics of $P_0(n)$ and $P_1(n)$ for measured $\overline{n}_0$.

\section{Unconditional accumulation of nonclassicality}

\begin{figure*}[!t]
\begin{center}
\includegraphics[width=2.1\columnwidth]{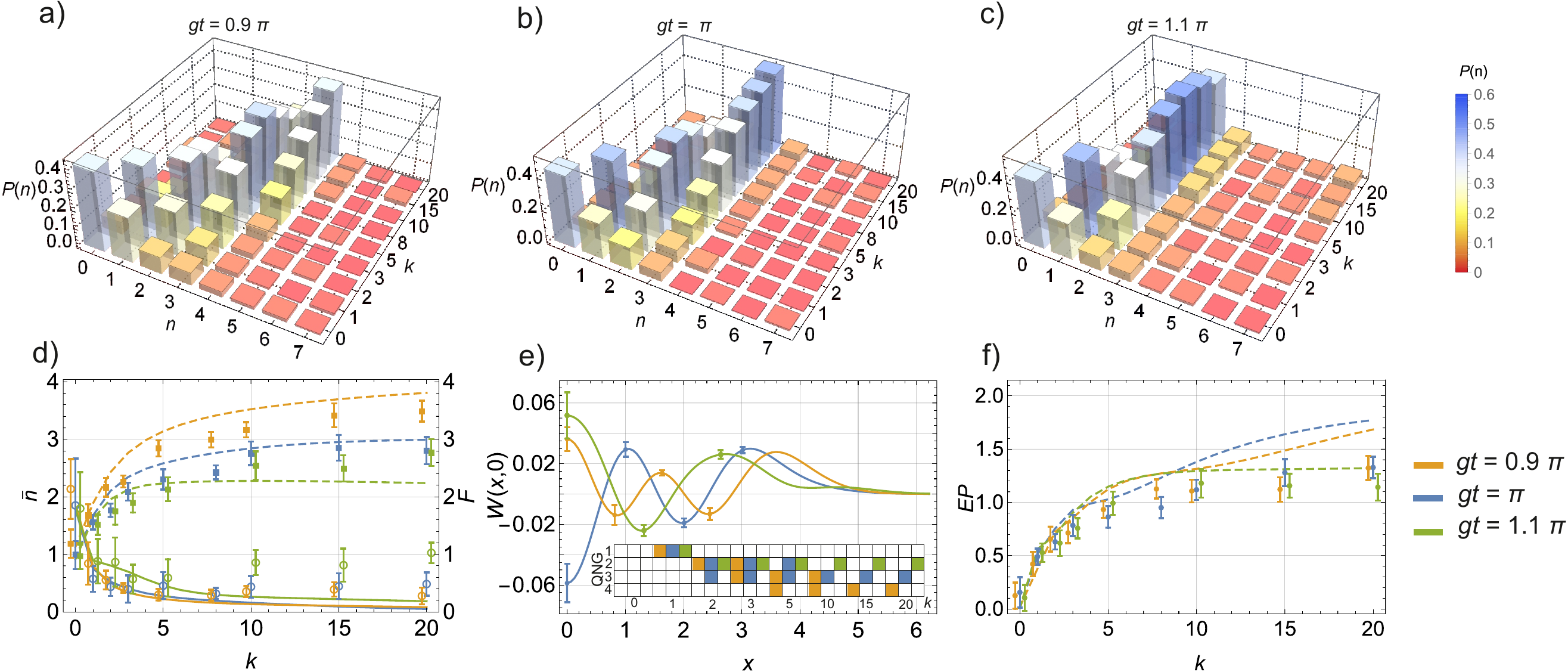}
\caption{Measurement results for nonclassicality accumulation by repetitive application of the nonlinear AJC coupling
 with number of repetitions  $k=0$ to 20 and for and the initial state of the atomic oscillator prepared in a thermal state with a mean phonon number $\overline{n}_{\rm th}=1.19\pm 0.04$. Figures a) to c) show the reconstructed output phonon number distributions $P_{\rm k}(n)$ for selected number of repetitions $k$ and for the interaction pulse area corresponding to $gt=0.9\pi, \pi$, and  $1.1\pi$, respectively. The graph in d) shows the evolution of the mean phonon number $\overline{n}$ and Fano factor $F$ with the dashed and solid lines corresponding to theoretical simulations and squares and circles displaying the measured data, respectively. Graph e) depicts the section of the Wigner function $W(x,0)$ evaluated from the measured $P_{\rm k}(n)$ for $k=20$ with inset showing the evolution in a hierarchy of the quantum non-Gausianity criteria~\cite{straka2018quantum}. The painted squares indicate the violation of non-Gaussianity criteria with each color assigned to the particular pulse area. The graph f) displays the evaluated entanglement potential $EP$ with clear monotonic increase with number of repetitions of the AJC interaction. The curves in d) - f) are theoretical simulations with no free fitting parameters and all error bars correspond to three standard deviations. We note that the residual quantitative differences between the measured and simulated values of nonclassicality criteria shown in graphs d) and f) are dominantly corresponding to small offsets in settings of the pulse areas $gt$ on the order of a few percent. The horizontal shift of the displayed $k$-values around the integer values $k$ has been introduced to avoid overlapping of the data points.
}
\label{fig:multipleAd}
\end{center}
\end{figure*}
Even a single operation~(\ref{eq:singleStepIdeal}) can significantly transform the state of a mechanical oscillator, a repetition of the same interaction may significantly increase the observable nonclassicality and quantum non-Gaussianity~\cite{blatt1995trapping,wallentowitz2004robust}. During the repetitions the operation remains constant; there is no optimization qubit state, interaction strength $g$, time duration $t$, or any other interaction characteristics. The process bears similarities to evolution of light inside a single resonant cavity with a nonlinear medium where nonclassicality is enhanced by a cyclic nonlinear interaction of the resonant optical mode with an off-resonant pump~\cite{bachor2004guide}. However, the accumulation has to not only overcome initial thermal occupation but also a heating caused by the resetting of internal population to the state $|S\rangle$.
The key aspect of the considered AJC~(\ref{eq:singleStepIdeal}) that is responsible for its properties is the nonclassical nonlinear modulation of all terms $P_{\rm k}(n)$ of the oscillator phonon population. This effect is independent on the initial population of the motional state and is enhanced by repeated application. It can no longer be understood just as repeated addition of single phonon, because the harmonic modulation terms of (\ref{eq:singleStepIdeal})  are merged and this leads to constructive enhancement of nonclassical aspects of output phonon populations $P_{\rm k}(n)$ which can be observed by several important metrics of nonclassicality.

Fig.~\ref{fig:multipleAd} shows the results of reconstruction of output phonon number distributions $P_k(n)$ for of up to $k=20$ repetitions of the AJC process for an initial thermal state with a mean phonon number~$\overline{n}=1.19\pm 0.04$.  They have been reconstructed from the measured Rabi oscillations after each displayed number of interaction repetitions $k$ for up to $n=7$. This dimension has been chosen so that the displayed $P_0(n)$ of the initial thermal state includes 99\,\% of its population. At the same time, this basis suffices for the observation of phenomena underlaying the accumulation process and allows for relatively small errors on $P_k(n)$ when employing solely measurements on first blue motional sideband, see Supporting material~\ref{subsec:processing} for more details. The initial thermal energy has been chosen so that it allows for clear illustration of important manifestations of accumulation dynamics for interaction~(\ref{eq:singleStepIdeal}), which include depopulation of motional ground state, population accumulation in phonon basis $|n\rangle$ corresponding to close to an integer multiplies of pulse area of $gt \sqrt{n+1}\sim 2 \pi$~\cite{blatt1995trapping}, and effect of heating corresponding to random photon recoils accompanying the reset of internal state.
The initial thermal population $P_0(n)$ is again transformed into a nonclassical phonon number distribution $P_1(n)$ after first step mainly due to the population shift from $P_0(0)$ to $P_1(1)$. However, its further repetitions accumulate other nonclassical aspects of modulation~(\ref{eq:singleStepIdeal}).


The first qualitative feature can be seen from the evolution of truncated phonon number distributions $P_k(n)$, which are depicted in Fig.~\ref{fig:multipleAd}-a)-c) for three different pulse areas $gt=0.9 \pi; \pi$ and $1.1 \pi$. We test these slightly different $gt$ to demonstrate the feasibility of control of the accumulation process.
We can see that in all scenarios the repetition of the procedure transforms the initial thermal state into a state closely resembling Fock state. The particular created Fock state depends on the pulse area and it is, respectively for the three pumping areas, $|4\rangle$, $|3\rangle$, and $|2\rangle$. This transition can be also seen from the monotonous increase of the states' average phonon number plotted in Fig.~\ref{fig:multipleAd}-d).

Insight into these features of nonclassicality accumulation can be gained by considering its asymptotic properties, similarly to \cite{slodivcka2016deterministic,marek2016deterministic}. When the pulse area satisfies condition
\begin{equation}\label{eq:cond}
gt\sqrt{ n+1} = 2 l\pi
\end{equation}
for any natural numbers $n$ and fixed $l$, sufficiently high number of perfect operations (\ref{eq:singleStepIdeal}) transforms any state with initial phonon number distribution $P_0(n)$ into an asymptotic  mixture of Fock states for large $k$:
\begin{equation}\label{}
    \rho_{\rm \infty } = \sum_{j=1}^{\infty} P_{\rm \infty }(n_j)|n_j\rangle\langle n_j|,
\end{equation}
where $n_j$ are the phonon numbers satisfying the condition~(\ref{eq:cond}) for given $l$. The probabilities of the mixture can be obtained from the initial phonon number distribution as $ P_{\rm \infty}(n_j) = \sum_{m = n_{j-1} +1}^{n_j}P_0(m)$ where, for the sake of notation, we assume $n_0 = -1$. For example, repeating operations with pulse area $gt = \pi$ presented in the Fig.~\ref{fig:multipleAd}-b) produces mixture with Fock states $n_j = 3, 15, 35, \ldots$, operations with pulse area $gt = 2\pi/\sqrt{3}$ produce mixture with $n_j = 2, 11, 47,\ldots$, and operations with pulse area $gt = 2\pi/\sqrt{5}$ mixture with $n_j = 4, 19, 79, \ldots$. This behavior also well explains the phonon number distributions in Fig.~\ref{fig:multipleAd}-a)~and~c), because the chosen pulse areas $gt = 0.9 \pi \sim 2\pi/\sqrt{5}$ and $gt = 1.1 \pi \sim 2\pi/\sqrt{5}$ are close enough to the theoretical values. The fit for $gt = 1.1 \pi$ is slightly worse which manifests as visible degradation of the Fock state for higher number of repetitions. This also demonstrates the need for high precision in practical setting of the pulse area $gt$.

The fundamental limitation on the achievable population of the Fock state $P_k(n_j)$ in the accumulation procedure is given by the sum of populations $\sum_{n=0}^{n_j} P_0(n)$ of the initial thermal state. However, the practical limit for Fock states with high $n_1$ will be mostly set by the requirement of high number of iterations $k$, the ability to control the applied pulse area $gt$ with a very high precision, and by the effective thermalization probability in the repumping process.
In the example presented in the Fig.~\ref{fig:multipleAd}-b) corresponding to $gt = \pi$, the population $P(3)$ reaches $P_{20}^{\rm exp}(3)=0.52\pm0.01$ after 20 accumulation steps. The discrepancy with the theoretical prediction $P_{20}^{\rm th}(3)=0.63$ evaluated from the model presented in Supplemental information~S1 can be attributed to a residual offset in an experimental setting of the pulse area $gt$. Its estimation from the fit of the measured photon number distribution $P_{20}(n)$ results in $gt=1.026 \pi$ corresponding to $P_{20}^{\rm th}(3)=0.54$, which is in a very good agreement with the measured value. While the fundamental limit on the achievable population $P(3)$ corresponding to the initial mean thermal phonon number $\overline{n}_{\rm th}=1.19\pm 0.04$ is $P_{\rm \infty}(n_1=3)\sim 0.91$, the measured value is further limited by the finite contrast of the applied $\pi$-pulse $\kappa = 97$\,\% and by the effective thermalization probability given by $\eta_{\rm eff}=0.17\pm 0.04$. The thermalization can't be fully avoided in most experimental scenarios, however, the close-to-ideal contrast $\kappa \sim 100$\,\% is feasible and would result in $P_{20}(3)=0.66$ after 40~repetition steps. A direct reduction of the thermalization rate could be achieved by utilization of higher trapping frequency with prospects of asymptotically reaching $P_{\rm \infty}^{\rm th}(3)=0.84$ for the presented experimental parameters and for $\omega_{\rm ax}\sim 2 \pi\times 5 $~MHz. Theoretical estimation with an ideal $\pi$-pulse contrast and no thermalization would result in $P_{20}(3)=0.88$, and reach $P_{43}(3)=0.91$ after reasonable 43 repetition steps.

Initial thermal state with higher energy leads to lower purity of the produced state. It can, however, lead to states with greater weight of higher Fock states. We can also see that for any energy of the input thermal state we can, in principle, design a pulse that eventually produces a Fock state with purity that is arbitrarily close to one. Such pulse would have low area $gt$ that would need to be set with very high precision. In theory, a laser pulse with an area $gt = 2\pi/\sqrt{21}$ would asymptotically produce Fock state $|20\rangle$ with element $P_{\rm \infty}(20) >0.9$ for any input thermal state with $\langle n \rangle < 9$.

The highly quantum non-Gaussian aspects of the states resulting from the accumulation process can be further evidenced in the reconstructed Wigner function $W(x,0)$, which has been evaluated from the measured $P_k(n)$ as an incoherent sum of Wigner quasi-distributions functions for $|n\rangle$ corresponding to state $\rho_{\rm k}=\sum_{n=1}^{7} P_{\rm k}(n) |n\rangle\langle n|$. The resulting Wigner functions effectively illustrate the state with population $P_k(n)$ and with randomized phase, which can be always implemented by random phase shift of the local oscillator in the reconstruction process. The corresponding data shown in the Fig.~\ref{fig:multipleAd}-e) point to several crucial aspects of the phonon distributions resulting from the accumulation process. The $P_{\rm k}(n)$ converges to distribution with two unambiguously negative concentric annuli in phase space which are directly observable, i.e without any correction for noise contribution or phonon detection efficiency. The number of observable negative regions increases correspondingly as the phonon population traverses to higher phonon numbers and, at the same time, concentrates in particular number state $|n\rangle$. This is further manifested a graduate transition in a faithful hierarchy of non-Gaussianity criteria~\cite{straka2018quantum} up to 4~phonons in increasing order shown in the inset. It shows that, indeed, accumulated phonon number states and their quantum non-Gaussian aspects remain well limited to maximally $n$~phonons.

We use the measured phonon number distributions to evaluate entanglement potential (EP) of quantum non-Gaussian states for future applications. EP, which is defined as logarithmic negativity \cite{vidal2002computable,plenio2005logarithmic} of entangled state created from the studied state by energetically passive coupling between two mechanical oscillators~\cite{asboth2005computable,toyoda2015hong}, is plotted in Fig.~\ref{fig:multipleAd}-f). We can see that even though the increase in nonclassicality is best visible in the first step in which the oscillator state goes from vacuum to mostly single phonon state, it still monotonously increases with the number of repetitions.
Importantly, this effect can be seen even though the measurements include a random and unavoidable diffusion of phonon number statistics due to the excitation and decay on the reshuffling transition with finite Lamb-Dicke parameters. The accumulation process is apparently robust against experimental imperfections and can be applied also to states with high thermal energy resulting from a simple Doppler cooling process, irrespectively to additional heating caused by the reseting of electronic state.



\section{Conclusions}

We have experimentally verified that nonclassicality of the generated phonon number distributions can be unconditionally accumulated. It is achievable by the modulation of thermal phonon number distribution using a natural Rabi interaction with a two-level system~\cite{blatt1995trapping,davidovich1996quantum,wallentowitz2004robust}. It represents a highly nonlinear extension of nonclassicality accumulation from single-resonant optical parametric oscillators to the platforms which allow for a direct implementation of Rabi interaction~\cite{xiang2013hybrid,leibfried2003quantum,raimond2001manipulating,fluhmann2019encoding,campagne2019stabilized,niemczyk2010circuit,lo2015spin,todorov2010ultrastrong,hartke2018microwave,kounalakis2019synthesizing}. The realized experiment demonstrates an unprecedented possibility of deterministic generation of quantum non-Gausian properties for controllable nonlinear interactions and promises a feasible bypass for no-go theorems for Fock state processing~\cite{berry2010linear,berry2011preservation}. The presented nonlinear interaction can be directly extended to nonlinear couplings in a solid-state mechanical oscillators~\cite{golter2016optomechanical,chu2018creation,sohn2018controlling,gieseler2019single} and generation of nonclassicality in experimental systems of several coupled oscillators and spins~\cite{oeckinghaus2019spin,rabl2010quantum,lee2017topical}.

This work has been supported by the grant No.~GA19-14988S of the Czech Science Foundation, CZ.02.1.01/0.0/0.0/16\_026/0008460 of MEYS CR and Palacky University IGA-PrF-2019-010. R.F. also acknowledges national funding from the MEYS and the funding from European Union's Horizon 2020 (2014-2020) research and innovation framework programme under grant agreement No. 731473 (project 8C18003 TheBlinQC, QuantERA ERA-NET Cofund in Quantum Technologies).

\clearpage

\newpage

\setcounter{equation}{0}
\setcounter{figure}{0}
\setcounter{table}{0}
\setcounter{page}{1} \makeatletter
\renewcommand{\theequation}{S\arabic{equation}}
\renewcommand{\thefigure}{S\arabic{figure}}
\renewcommand{\theHequation}{Supplement.\theequation}
\renewcommand{\theHfigure}{Supplement.\thefigure}
\renewcommand{\bibnumfmt}[1]{[S#1]}
\renewcommand{\citenumfont}[1]{S#1}

\section*{Supporting information: Unconditional accumulation of nonclassicality in a single-atom mechanical oscillator}

\section{Model of deterministic nonclassicality accumulation}
\label{subsec:model}

The process of accumulating nonclassicality corresponds to an unconditional repetition of nonlinear interaction, where the phonon number distribution after the $k$-th step $P_k(n)$ progressively change towards forms with stronger and stronger nonclassical properties~\cite{blatt1995trappingS}. The full model of the operation can be described as follows:
At the beginning of each of $k$ steps, the ion is prepared in separable state of the motional degree of freedom and the effective system of internal energy levels:
\begin{equation}
    \hat{\rho}_k \otimes |S\rangle\langle S|.
\end{equation}
The anti-Jaynes-Cummings interaction corresponding to~\cite{leibfried2003quantumS}
\begin{equation}
\hat{H}_{\rm blue}=g/2 (\hat{a}^{\dag} \hat{\sigma}_+ + \hat{a} \hat{\sigma}_-).
\label{eq:blueIntSupp}
\end{equation}
is effectively implemented with probability $\kappa$ transforming the joint system into:
\begin{equation}
   (1-\kappa) \hat{\rho}_k \otimes |S\rangle\langle S| + \kappa \left[ \hat{A} \hat{\rho}_k \hat{A}^{\dag} \otimes |S\rangle\langle S| + \hat{B} \rho_k \hat{B}^{\dag} \otimes |D\rangle\langle D|\right],
\end{equation}
where
\begin{equation}
    \hat{A} = \cos(gt \sqrt{\hat{n} + 1}),\quad \hat{B} = \hat{a}^{\dag} \frac{\sin(gt \sqrt{\hat{n}+1})}{\sqrt{\hat{n}+1}}.
\end{equation}
The electronic excited state $|D\rangle$ is then optically pumped back to the ground level $|g\rangle$. While doing this, the state of the motion suffers from slight thermalization due to the finite probabilities of photon recoil during the 854~nm photon absorbtion and 393~nm photon emission process, that can be expressed by the map
\begin{equation}\label{}
    \mathcal{D}(\hat{\rho_k}) = \int \frac{1}{2\pi\eta_{\rm eff}^2} e^{-\frac{|\alpha|^2}{\eta_{\rm eff}^2}} \hat{D}(\alpha)\hat{\rho_k}\hat{D}^{\dag}(\alpha) d^2\alpha.
\end{equation}
Operator $\hat{D}(\alpha) = \exp( \alpha \hat{a}^{\dag} - \alpha^{*}\hat{a})$ is the displacement operator and $\eta_{\rm eff}$ parameterizes the strength of the thermal fluctuations. When the thermal fluctuations are small, we can expand the exponential functions up to the second order, cancel the rapidly oscillating phase dependent terms, and effectively represent the mapping by
\begin{eqnarray}\label{approx}
\mathcal{D}(\rho)=\{\hat{\rho} + \eta_{\rm eff}^2 [ \hat{a}\hat{\rho}\hat{a}^{\dag} & +\hat{a}^\dag\hat{\rho}\hat{a}-  (\hat{n}+1/2)\hat{\rho} -  \nonumber\\
&- \hat{\rho}(\hat{n}+1/2)] + O(\eta_{\rm eff}^4)\}.
\label{eq:redistribution}
\end{eqnarray}
In our experiment, thermalization parameter $\eta_{\rm eff}$ effectively corresponds to the sum of squares of Lamb-Dicke parameters on spontaneous Raman transition $|D\rangle\rightarrow 4{\rm P}_{3/2}\rightarrow|S\rangle$. 

The full single step of the operation can be finally expressed as
\begin{equation}\label{}
    \hat{\rho}_{m+1} = (1-\kappa) \hat{\rho}_m + \kappa \left[ \hat{A} \hat{\rho}_m \hat{A}^{\dag}  + \mathcal{D}(\hat{B} \rho_m \hat{B}^{\dag})\right].
\label{eq:fullStepSupp}
\end{equation}

\section{Reconstruction of phonon number distributions}
\label{subsec:processing}

The motional phonon number distributions $P(n)$ of employed axial motion of a single trapped ion are obtained by probing and fitting of Rabi oscillation on the first blue motional sideband of at $|S\rangle =4{\rm S}_{1/2}(m=-1/2) \leftrightarrow |D\rangle = 3{\rm D}_{5/2}(m=-1/2)$ transition with 729~nm laser~\cite{leibfried2003quantumS}. For each length of the excitation laser pulse, exactly hundred repetitions of the excitation and internal state detection process are realized to minimize the effect of projection noise in the observable Rabi oscillation signal.

For the Lamb-Dicke parameter $\eta_{729}=\cos(\theta) k x_0 $, where $\theta$ is the the angle between the excitation 729~nm laser direction and motional axis, $k$ is the wavenumber corresponding to the 729~nm laser, and $x_0=\sqrt{\hbar/(2 m \omega_{\rm ax})}$ is the spread of the ion's position in the motional ground state, the Lamb-Dicke regime is defined by $\eta_{729}^2 (2 n+1)\ll 1$.  In this approximation, the measured oscillation can be described by
\begin{equation}
P_{\rm D}(t) = \sum_{n} P(n) \sin^2[\Omega_{n,n+1} t]e^{-\gamma(n) t},
\label{eq:bsb_flop}
\end{equation}
where $ \Omega_{n,n+1} = \Omega_{0} \eta_{729} \sqrt{n+1}$ is the effective Rabi frequency of the interaction with atom prepared in the state $|n\rangle$. We note, that for the phonon number reconstructions presented in this manuscript, the laser interaction with the ion in the highest motional state of $|n\rangle=7$ is still well within the Lamb-Dicke regime as $\eta_{729}^2 (2 n+1)=0.06$.
In practice, the oscillation signal is affected by laser amplitude noise and phase noise corresponding to laser frequency and magnetic field fluctuations, which cause the oscillation to decay. In a good approximation, this can be phenomenologically taken into an account by including a decay parameter $\gamma(n)=\gamma_0 (n+1)^\beta$, which is proportionally faster for phonon states causing oscillation with higher frequencies~\cite{leibfried2003quantumS}. Decay parameter $\gamma_0=0.32\pm 0.01$~kHz was found experimentally by fitting the Rabi oscillations on the blue motional sideband for an ion prepared in motional ground state $|0\rangle$ and $\beta=0.5\pm 0.2$ was estimated by comparing Rabi flops from several lowest Fock states ranging from $|0\rangle$ to $|6\rangle$. The measured oscillations on the blue sideband were fitted by Eq.~(\ref{eq:bsb_flop}), which corresponds to a direct reconstruction of phonon number distribution $P(n)$.

Uncertainty of the measured phonon number statistics arises mostly from the intrinsic detection noise corresponding effective inability of discrimination between $|S\rangle$ and $|D\rangle$ states in finite number realizations of the electron shelving process with binary output~\cite{itano1993quantumS}. This noise can be substantially suppressed by realization of large number $N$ of experiments for each point of measured Rabi oscillations. $N=100$ repetitions in our measurements still leaves a small but considerable projection uncertainty for each measurement point according to the relation:
\begin{equation}
\sigma(p) = \sqrt{\frac{p(1-p)}{N}}
\end{equation}
where $p$ corresponds to the probability of electronic population of the excited state $|D\rangle$.
We use solely this minimal fundamental uncertainty as an input for the realization of Monte-Carlo simulations of uncertainties of reconstructed phonon number distributions. We simulate 100 realizations of each Rabi oscillation points with uncertainty given by the $\sigma(p)$ and each simulated Rabi oscillation signal is independently fitted to obtain $P_k(n)$ using the equation~(\ref{eq:bsb_flop}). The uncertainty for the evaluated non-classicality witnesses is calculated in the same way.

\section{Characterization of the measured nonclassicality}
\label{subsec:criteria}
Quantum nonclassicality is defined as the inability to represent quantum states as statistical mixtures of coherent states. However, quantifying the amount of nonclassicality present in the quantum states is a more complicated task. Instead of using a single figure of merit we therefore employ several indicators in order to get a more complete picture.

Since the presented accumulation is expected to approach, at least approximatively, phonon number states, we can take advantage of figures of merit tailored specifically to recognizing this states. Fano factor
\begin{equation}\label{}
    F = \frac{\langle n^2\rangle - \langle n \rangle ^2}{\langle n\rangle}
\end{equation}
is larger than one for classical states and zero for phonon number states. Observing reduction of Fano factor, especially for states diagonal in the Fock state basis, is a clear indication of increasing nonclassicality.

Even more tailored are the Klyshko's hierarchic criteria of nonclassicality \cite{klyshko1996observableS}. They can be conveniently employed when assessing nonclassical properties manifested dominantly in the high population of particular Fock state, as they is sensitive to population difference in three neighboring states. For a chosen phonon number $n$, the Klyshko's criterion is defined as
\begin{equation}
K_{n} = n P_{n}^{2} - (n + 1) P_{n+1} P_{n-1}
\end{equation}
and it approaches value of $n$ for ideal Fock states.



The broadly employed and experimentally accessible indicator of nonclassicality and quantum non-Gaussianity is the value of Wigner function at the origin of the phase space, that can be defined as mean of the parity operator:
\begin{equation}
W(0,0) = \frac{2}{\pi} \sum_{n=0}^{n_{max}} (-1)^{n} P(n).
\end{equation}
The value can be directly and with high precision calculated from the estimated phonon number distribution. For odd quantum states the origin value of Wigner function is negative and thus serves as excellent indicator of nonclassicality. Negativity of Wigner function at the point of origin is a sufficient witness of nonclassicality and also quantum non-Gaussianity, but it fails to detect nonclassicality in many states, such as even number states. This can be remedied by looking for negative values in other regions of the phase space, which gives only a qualitative insight into the state's properties.

Quantum state is nonclassical when it cannot be represented as a mixture of coherent states. Similarly, quantum state is quantum non-Gaussian, when it cannot be expressed as arbitrary mixture of Gaussian states
\begin{equation}\label{}
    \hat{D}(\alpha)\hat{S}(r)|0\rangle,
\end{equation}
where $\hat{D}(\alpha)$ is again displacement operator and $\hat{S}(r) = \exp( -r^{*} \hat{a}^2 + r \hat{a}^{\dag 2})$ is Gaussian squeezing operator. One of the indicators for non-Gaussianity is again the negativity of Wigner function, but there other witnesses. In~\cite{straka2018quantumS} it was shown that for any Gaussian quantum state there is relationship between phonon number distribution elements $P(n)$ and $\tilde{P}_{n+1} = \sum_{j = n+1}^{\infty} P(j)$ that cannot be avoided. This relationship can be used to define a hierarchy of witnesses for individual values $n$. Each witness from the hierarchy can be used to verify, whether the population  of the $n$-th Fock state is compatible with Gaussian states.  Specifically, within the set of single mode Gaussian states given by (S13), for any state with a given value $\tilde{P}_{n+1}$, there is a maximal value that can be attained by $P(n)$. An experimentaly measured state that has, for the given value of $\tilde{P}_{n+1}$, value $P(n)$ larger than the Gaussian maximum is then necessarily non-Gaussian.

Finally, to have an overall quantifier, we employ the entanglement potential~\cite{asboth2005computableS} defined as the amount of entanglement contained in the state
\begin{equation}
\rho_{\rm ent}=e^{\pi/4 (a b^\dag −a^\dag b)} \rho_{\rm k} \otimes |0\rangle\langle 0|e^{- \pi/4( a b^\dag −a^\dag b)}
\end{equation}
where $a$ and $b$ are the annihilation operators for the first and the second oscillator mode, respectively. In quantum optics, this operation corresponds to splitting an optical mode on a balanced beam splitter. The entanglement potential measure reflects the fundamental inability to observe entanglement behind the beam-splitter if the state $\rho_{\rm in}$ at the input isn't nonclassical and takes further advantage of greater availability of measures of entanglement relative to measures of nonclassicality. In our case we quantify the entanglement using the straightforwardly computable logarithmic negativity~\cite{vidal2002computableS}
\begin{equation}
LN(\rho_{\rm ent}) = \log_2{\parallel \rho_{\rm ent}^{\rm PT}\parallel}
\end{equation}
Here PT denotes partial transposition and $\parallel A\parallel={\rm Tr}\sqrt{A^\dag A}$ corresponds to trace norm. This measure is not unique~\cite{vogel2014unifiedS,lee1991measureS,gehrke2012quantificationS}, however, it can be easily numerically evaluated even for high-dimensional non-Gaussian states.

\section{Unconditional generation of nonclassicality from thermal states}
\label{subsec:singleStep}

\begin{figure}[!t]
\begin{center}
\includegraphics[width=1\columnwidth]{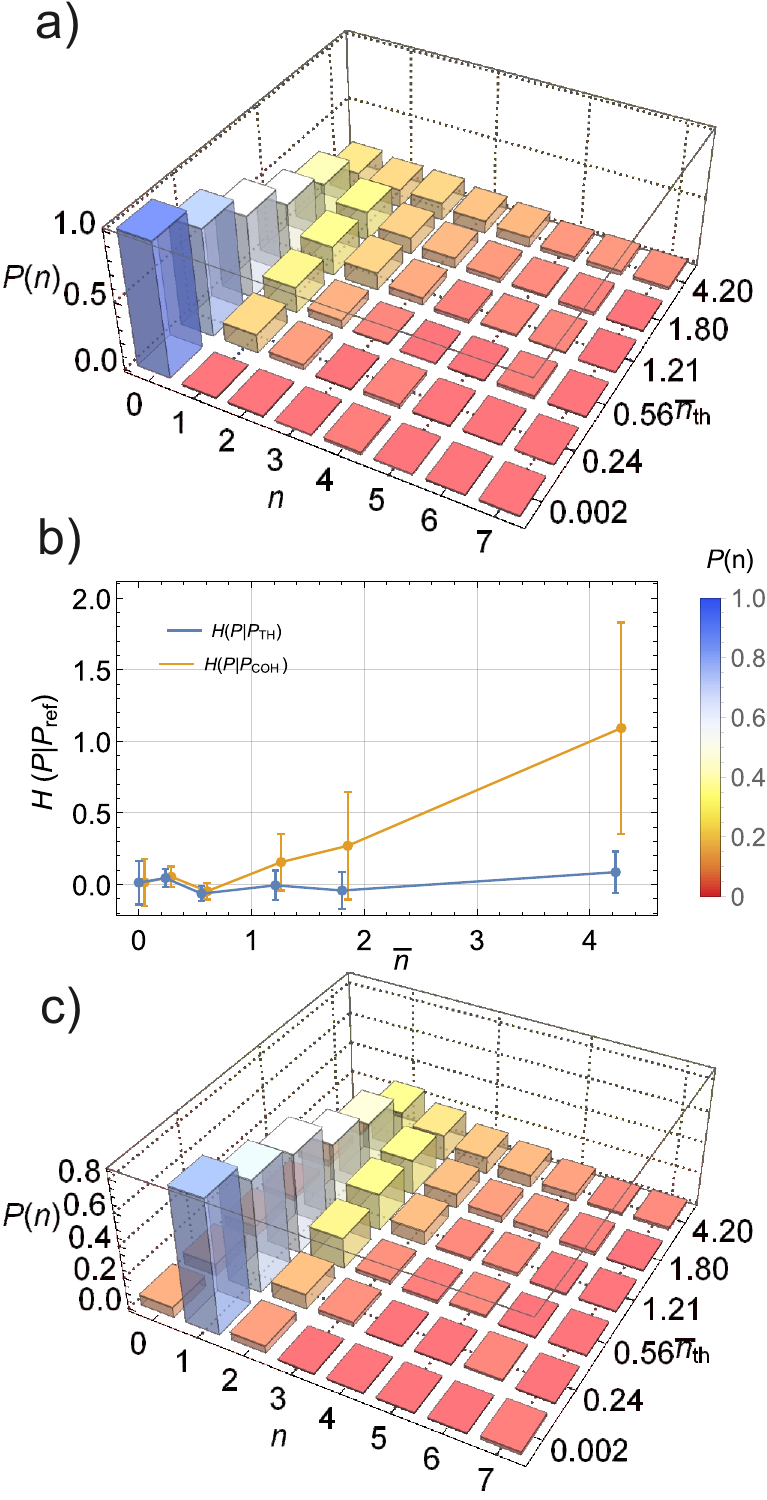}
\caption{The results of the realization of the single phonon emission. a) shows the initial phonon number distributions $P_0(n)$ for various mean phonon number values $\langle n \rangle$. The graph b) depicts the evaluated relative Shannon entropies for the measured phonon statistics and reference statistics $P_{\rm ref}(n)$ corresponding to ideal Bose-Einstein $P_{\rm TH}(n)$ and coherent states $P_{\rm COH}(n)$. The dimension of the considered Hilbert space is chosen to cover 99\,\% of the population of input thermal state with highest energy. The bar plot in c) shows the phonon number distributions $P_1(n)$ after single implementation of the nonlinear AJC interaction.}
\label{fig:singleFullData}
\end{center}
\end{figure}

The Fig.~\ref{fig:singleFullData}-a) and c) show the evaluated phonon number distributions $P_0(n)$ and $P_1(n)$ reconstructed from fits of measured Rabi oscillations on the blue motional sideband for input motional oscillator states and state of the motion after single step of implementation of AJC coupling with a pulse area of $gt=\pi$, respectively. The initial states after sideband cooling correspond very well to the states with close to ideal Bose-Einstein statistics within the errors evaluated using the Monte-Carlo simulations with input uncertainties corresponding solely to the projection noise in measured Rabi oscillations. To facilitate a quantitative comparison of the measured distributions to ideal Bose-Einstein statistics, we employ the fundamental definition of the thermally populated oscillator. It corresponds to realization with maximum Shannon entropy $H=-\sum_n p(n)\log p(n)$ for the given mean number of phonons. The evaluated entropy for the presented data is in agreement with the theoretically expected values within measurement errors. The distance between the ideal Bose-Einstein distribution $P_{\rm BE}(n)$ and the measured one can be accessed by the evaluation of the relative entropy $H(P(n)|P_{\rm BE}(n))$~\cite{Kul1951S}. The Fig.~\ref{fig:singleFullData}-b) shows the evaluated relative Shannon entropies for the measured statistics with respect to ideal Bose-Einstein and Poisson distributions corresponding to coherent states with the same mean phonon numbers. The relative proximity of the Shannon entropies to the states with ideal Bose-Einstein statistics with the same mean phonon number $\overline{n}$ is another confirmation of the classicality and thermal nature of the initial motional state.

\begin{figure}[!t]
\begin{center}
\includegraphics[width=0.9\columnwidth]{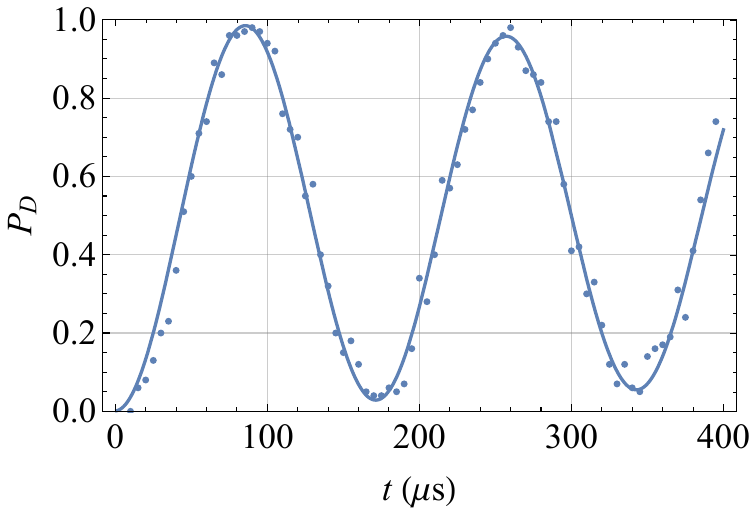}
\caption{Measured Rabi oscillations on the first blue motional sideband for the initial mean phonon population of $\overline{n}_{0}=0.002\pm 0.002$ corresponding to the $P_0(n)=0.995\pm 0.005$. The solid line corresponds to the simulation (\ref{eq:bsb_flop}) and the error bars were evaluated to one standard deviation of corresponding projection noise. }
\label{fig:flops}
\end{center}
\end{figure}
The AJC interaction with a pulse area $gt=\pi$ on the single ion prepared with an initial mean phonon number of $\overline{n}= (0.002 \pm 0.002)$ and $P_0(0)= (99.5\pm 0.5)$\,\% results in a single phonon addition with close to a $\kappa=97$~\% probability, which corresponds to the maximum of the first Rabi flop in the measurement presented in the Fig.~\ref{fig:flops}. However, optical pumping using the 854~nm reshuffling pulse is part or the experimental sequence and corresponds to the reset of internal population after each AJC interaction step. It leads to the redistribution of population from $P(1)$ to $P(0)$ and $P(2)$ with probabilities of 3.3\,\% and 6.6\,\%, respectively, corresponding to the effective thermalization factor $\eta_{\rm eff}=0.17\pm 0.04$. It has been estimated independently by measurement of the phonon number distribution resulting from the application of ten sequences consisting of reshuffling 854~nm pulse and carrier $\pi$~pulse on the initial motional state $|1\rangle$. In the presented measurements the polarization of the 854~nm beam propagating along the direction of the applied magnetic field has been optimized for the dominant depopulation of the $3{\rm D}_{5/2}(m=-1/2)$ level through the $\sigma^-$ transition to the $4{\rm P}_{3/2}(m=-3/2)$ state. In this way, the thermalization rate due to additional optical pumping on the 397~nm could be minimized while maintaining the possibility of efficient initial temperature control through sideband cooling process and utilization of high-fidelity Rabi flopping on the $4{\rm S}_{1/2}(m=-1/2)\leftrightarrow 3{\rm D}_{5/2}(m=-1/2)$ transition. The population of the Fock state $|1\rangle$ after complete single step~(\ref{eq:fullStepSupp}) corresponding to a nonlinear AJC interaction and reset of the atomic internal population is $P_1(1)=(84.6\pm 0.8)$\,\%. The Wigner quasiprobability distribution at the center of the phase space evaluated from the measured $P_1(n)$ gives $W(0,0)=-0.46\pm 0.01$.
We note that the effect of the thermalization through optical repumping could be also efficiently minimized for small mean phonon numbers by employing an additional $\pi$-pulse on the carrier transition, however, this would strongly compromise the general applicability of the presented scheme. As can be seen in the Fig.~\ref{fig:singleFullData}, the immediate manifestation and source of the nonclassicality in single step of AJC interaction corresponds to the flipping of population of motional ground state from $|0\rangle$ to $|1\rangle$. For higher initial thermal energies, the phonon distribution broadens and direct detection of nonclassicality becomes harder. The state $\rho_1$ is far from any Fock state squeezed in phonon-number distribution and exhibiting corresponding nonclassical interference in motional phase space. Nonclassical aspects are therefore disturbed by a tail of phonon number distribution for the first $\pi$-pulse. This increases demand on quality and repeatability of the measurement to identify such nonclassical phenomena.

\end{document}